**Magnetic domain dynamics in an insulating quantum ferromagnet**

D.M. Silevitch,[1] J. Xu,[2] C. Tang,[1] K.A. Dahmen,[3] T.F. Rosenbaum[1]

[1]Division of Physics, Mathematics, and Astronomy, California Institute of Technology, Pasadena, California 91125, USA
[2]The James Franck Institute, The University of Chicago, Chicago, Illinois 60637, USA
[3]Department of Physics, University of Illinois at Urbana-Champaign, Urbana, Illinois 61801, USA

ABSTRACT

The statistics and form of avalanches in a driven system reveal the nature of the underlying energy landscape and dynamics. In conventional metallic ferromagnets, eddy-current back action can dominate the dynamics. Here, we study Barkhausen noise in Li(Ho,Y)F$_4$, an insulating Ising ferromagnet that cannot sustain eddy currents. For large avalanches at temperatures approaching the Curie point, we find a symmetric response free of drag effects. In the low temperature limit, drag effects contribute to the dynamics, which we link to enhanced pinning from local random fields that are enabled by the microscopic dipole-coupled Hamiltonian (the Ising model in transverse field).

Since its initial observation in 1919 [1], Barkhausen noise has been a valuable tool for studying the dynamics of domain formation and motion in ferromagnets. The same basic physics – a change in macroscopic state via an ensemble of discrete microscopic jumps of widely varying size – extends across an eclectic variety of systems [2], including sheared foams [3], fluids in porous media [4], vortex avalanches in superconductors [5], magnetic Skyrmions [6], cascading disruptions of power grids [7], and even meme propagation on social media networks [8,9]. Statistical modeling of a distribution of such discrete events can be used to understand the energy scales, reversal mechanisms, and universality class underlying a particular physical system [2,10-12].

In metallic magnets, any abrupt change in the magnetization on both the bulk and microscopic levels is accompanied by a transient induced eddy current, which consequently produces a back-action magnetic field that acts as a drag effect on domain wall motion, hence slowing and skewing avalanche events [13-15]. While it is possible to minimize eddy current effects by studying thin films in which they are suppressed due to the geometry [16], they never can be completely eliminated in conductive magnetic materials. The drag effects from eddy currents hinder comparison to theory and also may mask the physics of other drag mechanisms. Here, we study Barkhausen noise events in an insulating ferromagnet, thereby precluding the confounding effects of eddy currents. Moreover, by choosing a material which is a realization of the simplest quantum magnet – an S=1/2 Ising spin system in transverse field – we are able to study the effects of random fields and, potentially, quantum fluctuations as a source of drag on domain reversal dynamics.

To study the behavior of magnetic avalanches in the absence of eddy-current drag, we turn to the insulating, dipolar-coupled Ising magnet, LiHo$_x$Y$_{1-x}$F$_4$ [17]. This family of rare-earth fluorides has band gaps in the infrared and visible [18,19], making them popular choices for lasing rods for near-IR lasers [20]. The Ho$^{3+}$ dipoles in the parent compound LiHoF$_4$ order ferromagnetically at a Curie temperature T$_C$ = 1.53 K [21]. The combination of the large band gap and the cryogenic temperatures needed to access the ordered magnetic state prohibits the generation of eddy currents during any magnetic avalanche event.

The holmium ions have a degenerate ground state doublet and a 9.4 K gap to the first excited spin state. At temperatures well below the gap energy, the material can be described by an effective S=1/2 Ising Hamiltonian with a saturation moment of 7 Bohr magnetons per spin [21-23]. Applying a magnetic field transverse to the Ising axis (H$_t$) mixes the ground state doublet with the first excited state, lifting the degeneracy and providing a mechanism for quantum tunneling between the two admixed, lowest-lying states. The effective Hamiltonian is then a realization of the Transverse Field Ising Model:

$$H = -\sum_{i,j} J_{ij} \sigma_i^z \sigma_j^z - \Gamma \sum_i \sigma_i^x \ , \qquad (1)$$

where $J_{ij}$ is the dipolar interspin coupling and $\Gamma \propto H_t^2$ is the tunneling term [24]. This Hamiltonian is further modified at temperatures below 0.4 K to include a rescaling of the electronic spin due to a hyperfine interaction with the Ho nuclear moments [24]. The combination of the long-range dipolar interaction and the tunable quantum tunneling via a laboratory field has made LiHoF$_4$ an important model system for studying transverse field-induced quantum phase transitions [25,26].

Chemical substitution of non-magnetic yttrium for holmium in LiHo$_x$Y$_{1-x}$F$_4$ permits tuning of the magnetic ground state from ferromagnet to spin glass to spin liquid with increasing *x* {Reich:1990ws}. Applying a transverse magnetic field to the dilution series similarly enables quantum tunneling and fluctuations. The combination of fluctuations and disorder permits tunable tunneling of domain walls [27], quantum annealing in complex free energy landscapes [28], and fluctuation-induced softening of domain pinning sites [29].

For *x* > 0.3, the system becomes a realization of the ferromagnetic Random-Field Ising Model (RFIM) with a Curie temperature suppressed linearly with *x* [29-32]. The effective electronic spin Hamiltonian for LiHo$_x$Y$_{1-x}$F$_4$ is:

$$H = -\sum_{ij} J_{ij} \sigma_i^z \sigma_j^z - \Gamma \sum_i \sigma_i^x - \sum_i h_{i,RF} \sigma_i^z \ , \qquad (2)$$

where $h_{i,RF}$ is a longitudinal field which varies randomly between sites [29]. Due to the off-diagonal components of the dipole-dipole coupling, $J_{ij}$, quantum fluctuations generated by internal transverse fields are present even in the absence of an externally-applied field. They play an important role in the domain reversal dynamics reported here by locally increasing pinning through random field effects and globally increasing the tunneling probability for a domain wall. The zero-field domain structure consists of micron-scale, needle-like domains oriented along the Ising (crystallographic *c*) direction [18,33]. The timescales associated with domain motion are experimentally accessible, with single-spin attempt frequencies measured to be of order 100 kHz in LiHo$_{0.44}$Y$_{0.56}$F$_4$ [27].

The experimental challenges associated with measuring Barkhausen-type behavior in LiHo$_x$Y$_{1-x}$F$_4$ arise from the temperature and magnetic field scales. To study the domain dynamics at temperatures well below the Curie temperature, sub-Kelvin cryogenics are required for even the parent compound. In addition, domain wall motion in such a quantum Ising system can involve the reversal of as few as 10 spins at once [27], resulting in very small signals in comparison to Barkhausen measurements on more conventional magnetic materials. We illustrate in Fig. 1a the apparatus developed to meet these challenges. A (5x5x10) mm$^3$ single crystal of LiHo$_{0.65}$Y$_{0.35}$F$_4$ was surrounded by thin sapphire plates to aid thermalization. A 150 turn pickup coil was closely wrapped around the center section of the crystal to detect Barkhausen events and eliminate edge effects. The static magnetization of the crystal was measured with a GaAs Hall magnetometer chip (Toshiba THS118). Both the pickup coil and the Hall magnetometer were wired in series with an identical empty set for background nulling.

This entire assembly was mounted on the copper cold finger of a helium dilution refrigerator equipped with an 8 T superconducting magnet oriented parallel to the Ising axis. The raw signal was boosted to detectable levels via a two-stage amplification technique. A superconductive-shielded high-frequency cryogenic transformer (CMR-Direct LTT-h), connected to the mixing chamber of the dilution refrigerator, amplified the differential pickup coil signal by a factor of 100. A battery-powered low noise preamplifier (Stanford Research SR560) outside the cryostat provided an additional 2000x gain. The amplified voltage was digitized with a 16-bit 250 kHz digitizer (National Instruments NI-6211). To minimize the effects of digital noise from the electronics, the entire cryostat and amplifier chain was located inside a Faraday cage with filtered power and ground, with the control and measurement computers located externally and communicating with the instrumentation via fiber optic bridges.

To acquire suitable statistics, we swept out hundreds of magnetic hysteresis loops spanning $\pm 8$ kOe, well above the $\pm 4$ kOe saturation field, at temperatures ranging from 80 to 700 mK (T$_C$ = 1.02 K for LiHo$_{0.65}$Y$_{0.35}$F$_4$ [30]). Typical hysteresis loops at 80 and 700 mK are shown in Fig. 1b. The field was ramped at rates ranging from 0.01 to 0.12 T/min (the maximum safe rate for this magnet); no difference was seen in the distribution or shapes of the observed events, indicating that the ramp rate was in the adiabatic limit for the sample. Only the data from the low-field region of the hysteresis loops was used for analysis, where the M vs. H curves are essentially linear ($\pm 1$kOe). An automated data analysis routine was used to search through the raw timeseries to identify and catalog the individual events. Events were identified by voltages more than 3.5$\sigma$ away from the mean after a low-frequency background subtraction with a 1 kHz cutoff, with the endpoints of each event determined by linear extrapolation back to the mean voltage (Fig. 1c). More complicated extrapolations led to qualitatively similar results.

While the predominant event behavior was single-signed as shown in Fig. 1c, a fraction of the longest duration events were preceded by an opposing-polarity precursor event. Such multi-domain reversals are illustrated in Fig. 1d and discussed below.

Before analyzing the behavior of the Barkhausen noise, we note first that the low-field behavior of the overall magnetization (Fig. 1b) is essentially temperature independent up to 700 mK, which reiterates the value of employing time-resolved measurements as a probe of the switching behavior and dynamics of this magnet.

Systems near criticality exhibit scaling relations and power-law behavior between various quantities in each system, and the exponents corresponding to these scaling relations help elucidate the underlying mechanisms and universality class for a particular system. For LiHo$_{0.65}$Y$_{0.35}$F$_4$ we plot histograms of the following quantities in Figures 2a and 2b with corresponding exponents: duration vs. event probability ($\alpha = 1.8 \pm 0.2$), area vs. probability ($\tau = 1.7 \pm 0.2$), energy vs. probability ($[(\tau - 1)/(2 - \sigma\nu z) + 1] = 1.7 \pm 0.1$), and finally in the voltage power-spectral-density ($\frac{1}{\sigma\nu z} = 1.7 \pm 0.1$). Exponents were extracted by fitting the data in the region that exhibited power-law behavior. We exclude the low and high tails of the data because of limitations in the instrumentation signal-to-noise, which artificially precludes small events from being detected, as well as a well-known phenomenological large-event cutoff due to demagnetization effects.

Although our exponents $\alpha$ and $\tau$ do not fall exactly onto the theoretical exponent values for either of the usual universality-classes seen in Barkhausen noise experiments ($\alpha \approx 2.0, \tau \approx 1.5$ for the long-range domain wall depinning universality class, and $\alpha \approx 1.5, \tau \approx 1.27$ for the short-range domain wall class), to within error bars, the exponent relationship $\frac{\alpha-1}{\tau-1} = \frac{1}{\sigma\nu z}$, characteristic of avalanches holds [10]. Finite size effects, small scaling regimes, disorder, and other effects are known to skew the measured values for the scaling exponents compared to the theoretical values extracted through finite size scaling and simulations of large systems [34]. The exponent values extracted here do in fact fall within the range of exponent values that have been quoted in the literature for other experiments that belong to the long range depinning universality class or the random field Ising model (RFIM) universality class for Barkhausen noise in disordered magnets [2], for which simulations [34] predict $\tau \approx 1.6$ and $\frac{1}{\sigma\nu z} \approx 1.75$, and $\alpha \approx 2$ and ($[(\tau - 1)/(2 - \sigma\nu z) + 1] \approx 1.42$). Future studies of the history dependence along the lines of Ref. [35] could help narrow down the underlying universality class further.

We plot in Fig. 2c the occurrence rate of avalanche events as a function of event area and duration, and fit each bivariate histogram to a power law. We find that the data is consistent with a power law exponent of 1 across the entire distribution at all T. It is likely that this exponent for the entire range includes strings of temporally overlapping avalanches, possibly through thermally triggered "aftershocks" of avalanches. Such overlaps of avalanches are known to produce size versus duration exponents around 1 [36]. Alternatively, we note that the event distribution at 80 mK appears to segregate into three visually-distinct regimes; if the power law fit is constrained to the central section (spanning 160 to 500 microseconds), we obtain a value of $\frac{1}{\sigma v z} = 1.5 \pm 0.1$ This suggests the possibility that the central regime may be the true avalanche scaling regime, but the limited range over which this behavior occurs prevents us from making a more definitive statement. The additional number of events associated with the crossover into this possible scaling regime is also visible in the duration probability histograms in Fig 2b as a small "bump" above the trend lines for the bins around 200 microseconds. Applying an external transverse field and thereby inducing quantum fluctuations potentially could broaden this postulated scaling regime, allowing for a more robust analysis.

As illustrated in Fig. 1d, a small fraction of the observed events are preceded by a smaller precursor feature of opposite sign. Some spins actually reverse against the direction of the local magnetic field immediately prior to the more typical domain reversal event where spins realign to point along the magnetic field. Such a counter-intuitive effect is possible due to the dipole nature of the interspin coupling in $LiHo_xY_{1-x}F_4$, where for some relative orientations the dipole coupling can be antiferromagnetic. Temporary reconfigurations of the spins to oppose the magnetic field can be energetically favorable locally, and hence enable a reverse-polarity switching event. Analogous behavior has been observed in plastic deformation of sheared materials, where local pinning sites similarly can distort the free energy landscape and allow for seemingly "uphill" motion. [37,38] Numerical simulations of domain-wall motion in disordered perpendicular-anisotropy thin films [39] likewise have observed a small fraction of reverse-polarity events. The occurrence probability of these precursor events is strongly temperature dependent, with approximately 5% of the 80 mK events showing a precursor, decreasing to 1% at 250 mK and less than 0.1% at 500 mK and above where thermal fluctuations dominate.

Domain dynamics in ferromagnets are often understood through the use of the ABBM model [40], which uses a mean-field framework to develop and exactly solve a Langevin equation for the velocity of the avalanches. One of the striking predictions of the ABBM model is that when ensemble averages of avalanche events are scaled by the event duration, the resulting lineshapes are symmetric about the center of the event [41]. In the absence of demagnetization effects, the scaled curve is an inverted parabola, with a flattening for finite demagnetization [16]. By contrast, measurements on many magnetic materials show scaled curves which skew to the left (earlier times), which in the ABBM framework implies a negative effective mass [14]. Any non-symmetric events must be due to some drag effect, but the physical origin of these drag effects, which have been observed in many different systems (fluid flow in disordered nanoporous materials[42], dislocation avalanches in metallic microcrystals [43], and creep rupture in heterogenous materials [44]) can vary from system to system.

In bulk metallic ferromagnets, the origin of these drag effects has been ascribed typically to eddy-current back-action. However, $LiHo_{0.65}Y_{0.35}F_4$, which is electrically insulating, cannot sustain eddy currents, so we must look beyond conductivity-related mechanisms in order to explain the observed drag. Furthermore, the typical modeling of drag in spin dynamics via the Landau-Lifshitz-Gilbert equation does not directly apply here due to the S=1/2 Ising nature of the spins, which precludes the spin-precession dynamics that underly the LLG approach. Direct coupling between the domain walls and phonons, another potential drag mechanism, is limited due to the low phonon density: the measurement temperatures are approximately three orders of magnitude lower than the Debye temperature $\theta_D$=560 K for the Li(Ho,Y)F$_4$ family [45]. However, it has been shown in previous experiments that domain-pinning arising from random fields, due to the quenched disorder from the dilution of Ho$^{3+}$ ions by Y$^{3+}$ ions, widens the hysteresis loop, which corresponds to greater dissipation of energy [29]. While the exact microscopic mechanism through which the random fields affect the spin-lattice coupling for LiHo$_{0.65}$Y$_{0.35}$F$_4$ is still an open question, the enhanced dissipation due to these random fields suggests that they are most likely involved in driving the observed drag effects.

We show in Fig. 3a and b scaled curves as a function of temperature for short (< 150 μs) and long (> 500 μs) time events, respectively. Following the approach in Ref. [14], we construct the scaled curves by first normalizing the time axis for each event by its individual duration, normalizing the voltage such that the total integrated area is equal to 1, and finally averaging the voltage as a function of normalized time across all events for a given temperature and duration bin. Despite the absence of eddy currents in insulating LiHo$_{0.65}$Y$_{0.35}$F$_4$, clear early-time asymmetries characterize a number of the curves in both time bins, indicating the presence of alternative sources for drag effects. We also include in Fig. 3b a fit of the 700 mK data to the ABBM model incorporating both demagnetization effects and a phenomenological linear skew term[14]. This ABBM-based model only accurately fits the observed data for long events at high temperature, with noticeably different shapes appearing in the short-time, low-temperature limit. These deviations point to a crossover to a different set of domain dynamics and consequent change In universality class away from mean-field [46,47].

The evolution of the asymmetry in the scaled lineshapes can be described quantitatively by calculating a normalized skewness [14] for the scaled lineshapes at each temperature/duration bin. For short events ($< 150~\mu$s), the skewness was essentially constant as a function of temperature, equal to $0.46 \pm 0.01$ for all temperatures between 80 and 700 mK. By contrast, for longer events ($> 500~\mu$s), the skewness decreases monotonically with increasing temperature, from 0.42 at 80 mK to a much more symmetric 0.18 at 700 mK. The overall trend is for longer events at high temperatures to tend towards lower skewness, similar to what has been observed in Barkhausen measurements on metallic ferromagnets [14,15], but opposite to the behavior of non-magnetic crackling noise systems such as yield transitions in sheared amorphous solids [48], granular media [49], and earthquakes [50].

The shorter duration events shown in Fig. 3a demonstrate only a weak dependence of average skewness on temperature. The long duration events of Fig. 3b, however, see approximately a factor of two decrease in skewness at 600 and 700 mK compared to the same duration bins at low temperature, indicating that drag effects are becoming proportionally less important in the dynamics of those events. While the short events correspond to individual domain reversals whose dynamics are dominated by the drag-inducing microscopic pinning landscape, the longer events represent a linked cascade of reversals more strongly dependent on the strength of thermal fluctuations, which overcome local pinning mechanisms.

Connecting the observed behavior with specific microscopic mechanisms is feasible due to the well-understood spin Hamiltonian of LiHo$_x$Y$_{1-x}$F$_4$. We note first that the drag effects in the short-time limit persist to well above 400 mK, the scale of the electro-nuclear hyperfine interaction, indicating that interactions with nuclear spins [25] can be ignored and hence the purely electronic Hamiltonian, given above in Eq. (2), is the applicable description. The quantum fluctuations arising from the second term in Eq. (2) impact the domain dynamics via a weakening of the interdomain walls, allowing for a soft and dissipative reversal process. By contrast, the random field effects in the third term in the Hamiltonian act to enhance pinning by locally aligning antiparallel to the external applied field and retarding the reversal process.

We elucidate the relative importance of these two effects by comparing the magnetic field scales for each effect. Previous experiments on the more dilute compound LiHo$_{0.44}$Y$_{0.56}$F$_4$ have shown that random field effects dominate at low transverse fields, with a crossover between the random-field enhanced pinning and quantum speedup due to enhanced tunneling occurring at 3.25 kOe [29]. However, the internal transverse fields generated by nearest-neighbor pairs of spins are at most a few hundred Oe [51] – far below the crossover field scale. Hence, in the absence of an externally-applied transverse field, random field pinning is the most likely mechanism for inducing drag in domain-wall motion. In this regime, short-duration events are pinned locally and are essentially robust against thermal fluctuations. For long-time avalanche events, by contrast, thermal fluctuations can wash out the weakest links in the random-field pinning landscape, starting a thermally-driven cascade. As the temperature approaches the Curie point and thermal fluctuations are amplified, an avalanche can proceed with no substantive drag.

We have examined Barkhausen noise events in a model, insulating quantum ferromagnet. By eliminating electronic conductivity, we are able to eliminate eddy currents, the most common source for drag effects in magnetic domain dynamics. We find that the ensemble average over long-duration avalanche events just below the Curie temperature approaches a symmetric lineshape, as expected for a drag-free response. This limit has not been observed in metallic ferromagnets. At lower temperatures and for short duration events, we do observe clear drag effects, which cannot arise from eddy currents. By studying the temperature dependence in different time regimes and comparing to the well-understood microscopic Hamiltonian of this model Ising magnet, we are able to propose the presence of two coexisting drag mechanisms: one arising from quantum-fluctuation-induced domain wall broadening and the other from random-field-induced domain wall pinning.

The disparate nature of the drag mechanisms discussed here can be probed in more detail by exploiting one of the key features of the Li(Ho,Y)F$_4$ family, the ability to tune the strength of both the quantum fluctuations [24] and the random fields [30] via application of an external magnetic field applied transverse to the Ising axis. The relative importance of the two effects can be controlled via the transverse field, the temperature, and the degree of yttrium substitution [29]. Continuously tuning the drag behavior within a single system offers a valuable experimental substrate for comparing classical and quantum Barkhausen noise.

We are grateful to P.C.E. Stamp for valuable discussions. The work at the California Institute of Technology was supported by US Department of Energy Basic Energy Sciences Award DE-SC0014866.

# References

[1] H. Barkhausen, Z. Phys. **20**, 401 (1919).
[2] J. P. Sethna, K. A. Dahmen, and C. R. Myers, Nature **410**, 242 (2001).
[3] S. Tewari, D. Schiemann, D. J. Durian, C. M. Knobler, S. A. Langer, and A. J. Liu, Phys. Rev. E **60**, 4385 (1999).
[4] M. Cieplak and M. O. Robbins, Phys. Rev. Lett. **60**, 2042 (1988).
[5] S. Field, J. Witt, F. Nori, and X. Ling, Phys. Rev. Lett. **74**, 1206 (1995).
[6] S. A. Díaz, C. Reichhardt, D. P. Arovas, A. Saxena, and C. J. O. Reichhardt, Phys. Rev. Lett. **120**, 117203 (2018).
[7] M. L. Sachtjen, B. A. Carreras, and V. E. Lynch, Phys. Rev. E **61**, 4877 (2000).
[8] J. P. Gleeson, J. A. Ward, K. P. O'Sullivan, and W. T. Lee, Phys. Rev. Lett. **112**, 048701 (2014).
[9] J. P. Gleeson and R. Durrett, Nat. Comms. **8**, (2017).
[10] G. Durin and S. Zapperi, Phys. Rev. Lett. **84**, 4705 (2000).
[11] D. V. Denisov, K. A. Lőrincz, W. J. Wright, T. C. Hufnagel, A. Nawano, X. Gu, J. T. Uhl, K. A. Dahmen, and P. Schall, Sci. Rep. **7**, 43376 (2017).
[12] F. Bohn, G. Durin, M. A. Corrêa, N. R. Machado, R. D. Della Pace, C. Chesman, and R. L. Sommer, arXiv 1801.09948 (2018).
[13] A. P. Mehta, A. C. Mills, K. A. Dahmen, and J. P. Sethna, Phys. Rev. E **65**, (2002).
[14] S. Zapperi, C. Castellano, F. Colaiori, and G. Durin, Nature Physics **1**, 46 (2005).
[15] F. Colaiori, G. Durin, and S. Zapperi, Phys. Rev. B **76**, 224416 (2007).
[16] S. Papanikolaou, F. Bohn, R. L. Sommer, G. Durin, S. Zapperi, and J. P. Sethna, Nature Physics **7**, 316 (2011).
[17] D. H. Reich, B. Ellman, J. Yang, T. F. Rosenbaum, G. Aeppli, and D. P. Belanger, Phys. Rev. B **42**, 4631 (1990).
[18] J E Battison, A Kasten, M J M Leask, J B Lowry, and B M Wanklyn, Journal of Physics C: Solid State Physics **8**, 4089 (1975).
[19] S. A. Payne, L. L. Chase, L. K. Smith, W. L. Kway, and W. F. Krupke, IEEE J. Quantum Electron. **28**, 2619 (1992).
[20] A. Dergachev, P. F. Moulton, and T. E. Drake, Advanced Solid-State Photonics **98**, 608 (2005).
[21] P. Hansen, T. Johansson, and R. Nevald, Phys. Rev. B **12**, 5315 (1975).
[22] P. B. Chakraborty, P. Henelius, H. Kjonsberg, A. W. Sandvik, and S. M. Girvin, Phys. Rev. B **70**, 144411 (2004).
[23] M. Schechter and P. C. E. Stamp, Phys. Rev. B **78**, 054438 (2008).
[24] D. Bitko, T. F. Rosenbaum, and G. Aeppli, Phys. Rev. Lett. **77**, 940 (1996).
[25] H. M. Rønnow, R. Parthasarathy, J. Jensen, G. Aeppli, T. F. Rosenbaum, and D. F. McMorrow, Science **308**, 389 (2005).
[26] H. M. Rønnow, J. Jensen, R. Parthasarathy, G. Aeppli, T. F. Rosenbaum, D. F. McMorrow, and C. Kraemer, Phys. Rev. B **75**, 054426 (2007).
[27] J. Brooke, T. F. Rosenbaum, and G. Aeppli, Nature **413**, 610 (2001).
[28] J. Brooke, D. Bitko, T. F. Rosenbaum, and G. Aeppli, Science **284**, 779 (1999).

[29]	D. M. Silevitch, G. Aeppli, and T. F. Rosenbaum, Proc. Natl. Acad. Sci **107**, 2797 (2010).
[30]	D. M. Silevitch, D. Bitko, J. Brooke, S. Ghosh, G. Aeppli, and T. F. Rosenbaum, Nature **448**, 567 (2007).
[31]	M. Schechter, Phys. Rev. B **77**, 020401 (2008).
[32]	S. M. A. Tabei, F. Vernay, and M. J. P. Gingras, Phys. Rev. B **77**, 014432 (2008).
[33]	A. Biltmo and P. Henelius, Epl **87**, 27007 (2009).
[34]	O. Perković, K. A. Dahmen, and J. P. Sethna, Phys. Rev. B **59**, 6106 (1999).
[35]	J. H. Carpenter, K. A. Dahmen, A. Mills, M. Weissman, A. Berger, and O. Hellwig, Phys. Rev. B **72**, 052410 (2005).
[36]	R. A. White and K. A. Dahmen, Phys. Rev. Lett. **91**, (2003).
[37]	M. Talamali, V. Petäjä, D. Vandembroucq, and S. Roux, Phys. Rev. E **84**, 016115 (2011).
[38]	S. Papanikolaou, Y. Cui, and N. Ghoniem, Modelling Simul. Mater. Sci. Eng. **26**, 013001 (2018).
[39]	T. Herranen and L. Laurson, Phys. Rev. Lett. **122**, 117205 (2019).
[40]	B. Alessandro, C. Beatrice, G. Bertotti, and A. Montorsi, J Appl. Phys. **68**, 2901 (1990).
[41]	F. Colaiori, Advances in Physics **57**, 287 (2008).
[42]	X. Clotet, S. Santucci, and J. Ortín, Phys. Rev. E **93**, 012150 (2016).
[43]	G. Sparks and R. Maaß, Acta Materialia **152**, 86 (2018).
[44]	Z. Danku and F. Kun, Phys. Rev. Lett. **111**, (2013).
[45]	R. L. Aggarwal, D. J. Ripin, J. R. Ochoa, and T. Y. Fan, J Appl. Phys. **98**, 103514 (2005).
[46]	L. Laurson, X. Illa, S. Santucci, K. T. Tallakstad, K. J. Måløy, and M. J. Alava, Nat. Comms. **4**, 2927 (2013).
[47]	A. Dobrinevski, P. Le Doussal, and K. J. Wiese, Epl **108**, 66002 (2014).
[48]	C. Liu, E. E. Ferrero, F. Puosi, J.-L. Barrat, and K. Martens, Phys. Rev. Lett. **116**, 065501 (2016).
[49]	J. Barés, D. Wang, D. Wang, T. Bertrand, C. S. O'Hern, and R. P. Behringer, Phys. Rev. E **96**, 052902 (2017).
[50]	A. P. Mehta, K. A. Dahmen, and Y. Ben-Zion, Phys. Rev. E **73**, 056104 (2006).
[51]	C. M. S. Gannarelli, D. M. Silevitch, T. F. Rosenbaum, G. Aeppli, and A. J. Fisher, Phys. Rev. B **86**, 014420 (2012).

**Figure Captions**

1. Measurement of Barkhausen events in LiHo$_{0.44}$Y$_{0.56}$F$_4$. (a) Schematic of the experimental setup. A pickup coil is wound around a single-crystal sample (inset) and placed on the cold finger of a helium dilution refrigerator inside a superconducting solenoid magnet. The induced voltages are amplified at low temperature by a high-frequency transformer, at room temperature by a low-noise voltage preamp, and then digitized to a computer. (b) Typical magnetization hysteresis loops, measured using a GaAs Hall magnetometer, at T=80 and 700 mK. The overall behavior of the magnetization in the low-field regime (|H|<1 kOe) is essentially temperature-independent (c) Voltage time series showing an individual, single-polarity Barkhausen event. (d) Voltage time series showing a long-duration event with an opposing-polarity precursor stage.
2. Statistical distribution of events for a series of temperatures. (a) Probability distribution of event area (integral of V*time) and energy (V^2 * time). Power laws are close to mean-field predictions. (b) Probability distribution of event duration and power spectral density of events (c) Bivariate histograms of event probability vs. area and duration at T=80 mK and 600 mK, showing an approximately linear relationship with no well-defined crossover. Lines are best-fit power laws (offset for clarity). For events with opposing-polarity precursors (Fig 1c), only the forward-polarity portion is included in the histogram analyses. Since these precursors are both rare and considerably smaller than the associated forward event, incorporating them into the histograms would have minimal effect.
3. Evolution of lineshape as a function of temperature and event duration, showing the onset of drag effects. (a,b) Average scaled event shape vs. scaled duration (see text for details on averaging and scaling process) for short (<150 μs) and long (>500 μs) events, respectively. Dashed line is a fit at T = 700 mK to the ABBM model incorporating shape demagnetization and a phenomenological skew term. [14]

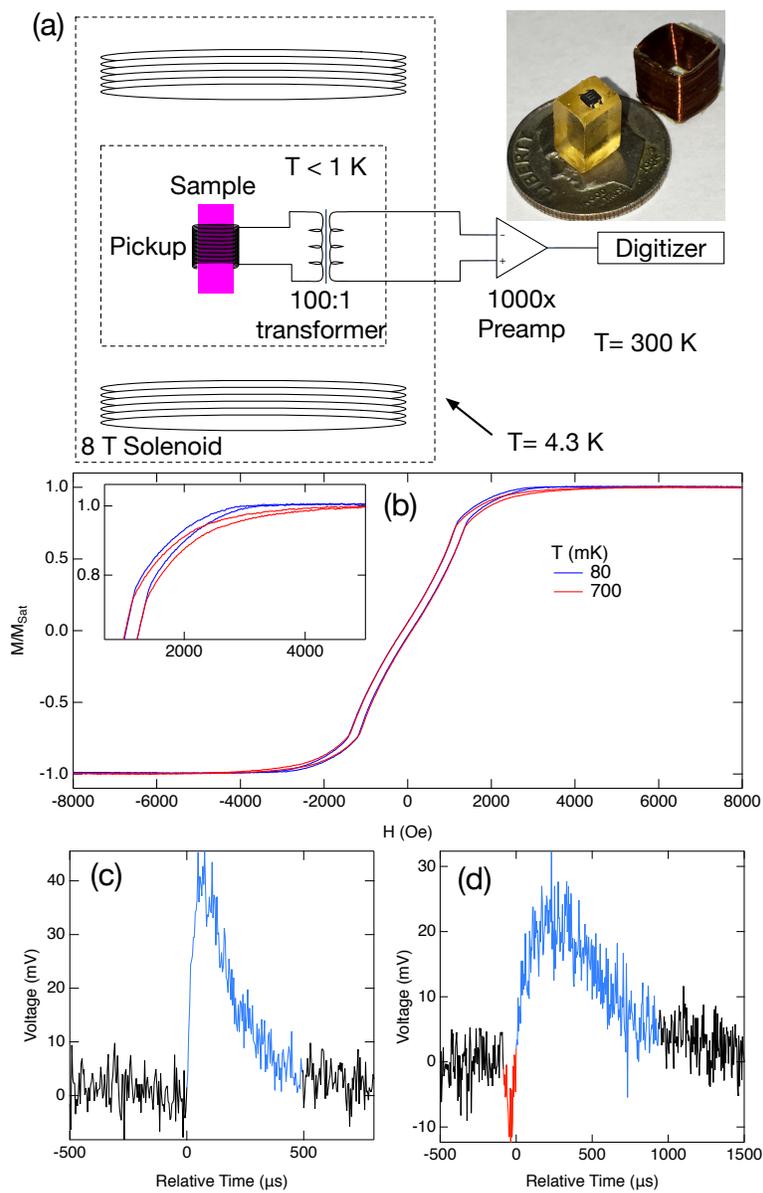

Figure 1

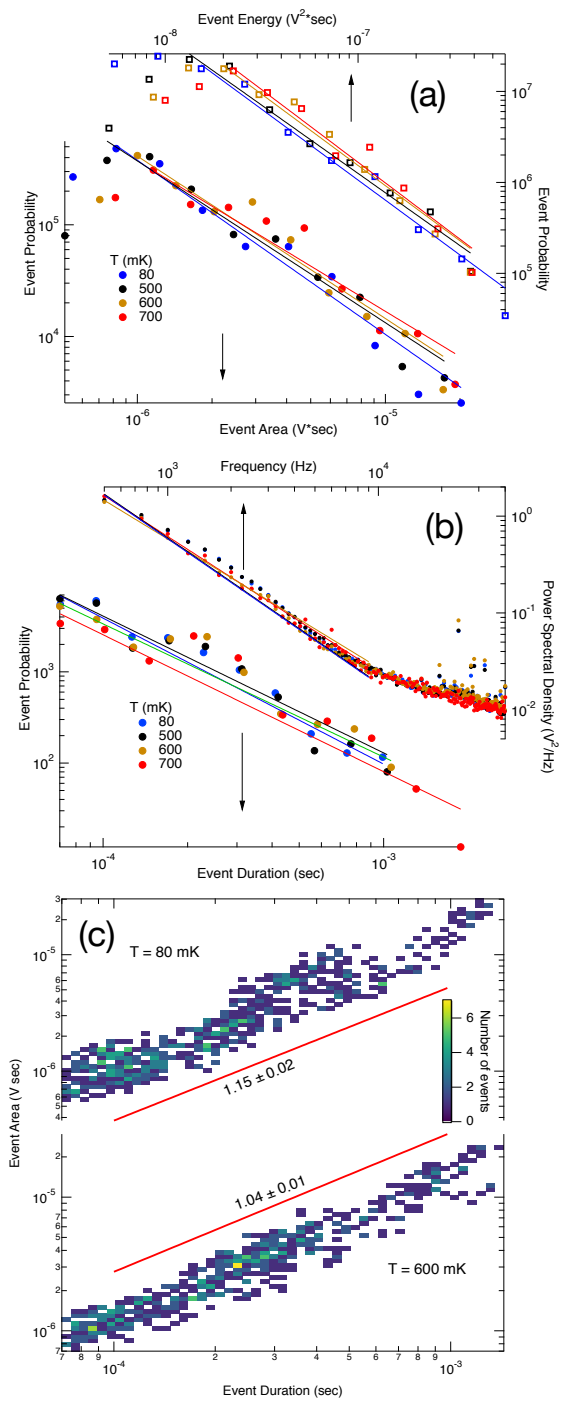

Figure 2

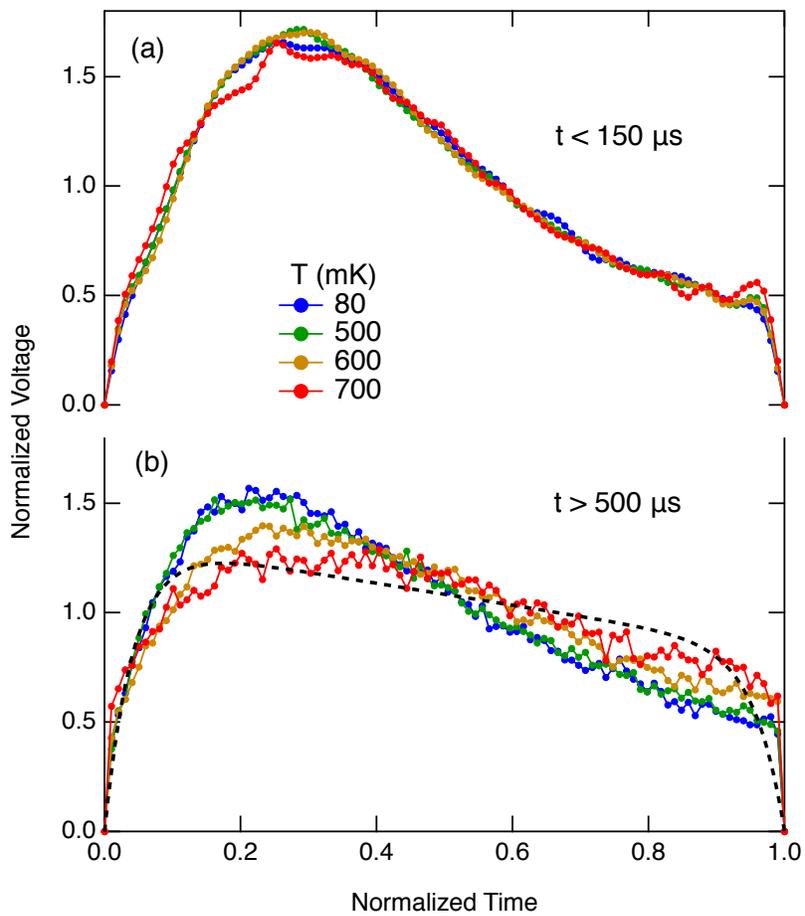

Figure 3